\documentclass[prc,aps,manuscript]{revtex4}
\usepackage{graphicx}

\begin{document}
\draft
\title{Splashing and evaporation of nucleons from excited nuclei}
\author{V.M. Kolomietz \footnote{Permanent address: Institute for Nuclear Research, Prospect Nauki 47,\ 03680
Kiev, Ukraine \newline e-mail: vkolom@ph.tum.de
\newline  e-mail: vkolom@kinr.kiev.ua }}
\address{Physik-Department TU M\"{u}nchen, 85747 Garching, Germany}

\begin{abstract}
The energy spectrum and the emission rate of particles emitted from excited
nucleus due to both the evaporation and the splashing (emission from a cold
vibrating nucleus) are calculated. We show that the collective motion of the
nuclear Fermi liquid is accompanied by direct non-statistical emission of
nucleons via the dynamical distortion of the Fermi surface.

\bigskip

\bigskip


\bigskip

\bigskip


\end{abstract}

\maketitle

\section{Introduction}

The emission of particles from a collectively excited state of the
nucleus as a Fermi liquid drop can occur in two ways. First, due
to the relaxation processes the collective energy is transferred
to the intrinsic degrees of freedom with subsequent evaporation of
particles. On the other hand, a direct non-statistical emission
(splashing) of nucleons is also possible via the dynamical
distortion of the Fermi surface accompanying the collective
motion. In general the relative contributions of these mechanisms
depend upon the magnitude of the nuclear friction coefficient. In
this work, the limiting cases of the direct (non-statistical)
particle emission from the non-damped giant multipole resonance
(GMR) and the particle evaporation from the heated nucleus are
compared. The direct particle emission from the GMR has been
extensively studied within pure quantum mechanical approaches
(see, e.g. \cite{stva79,woud87,woud90,bobr89,bobr90,muur94}). We
exploit here a semiclassical phase-space theory to study the
energy spectra and the escape width for nucleons directly emitted
from a collective state (splash out effect). Within this approach,
the particle emission is describing in a more qualitative way than
in microscopic quantum mechanical models. However an advantage of
this approach is the possibility to take into account the
statistical, direct and cascade emissions of the particles from
the heated nucleus which is simultaneously involved in the
collective motion. In particular, this approach can be used to
describe both the evaporation of neutrons and the direct
(non-statistical) neutron emission at the nuclear descent from the
fission barrier to the scission point. Below, we will restrict
ourselves to the case of particle emission from the giant monopole
resonance. A generalization of our approach to the case of an
arbitrary multipole giant resonances can be done in a
straightforward way.

\section{Emission of particles caused by the Fermi-surface distortion}

For an excited nucleus, the particle emission rate can be calculated from
the assumption that the nucleons localized in a single-particle mean field
of depth $V_{0}$ and described by the phase-space distribution function $f(%
\vec{r},\vec{p},t)$ escape out the nuclear surface if the energy of their
radial motion exceeds $V_{0}$. The flux density of neutrons emitted from the
nuclear surface is given by, see also \cite{FrLy},
\begin{equation}
J_{{\rm emis}}(t)=\left. \Theta (E)\int {\frac{g_{s}\,d \vec{q}}
{(2\pi \hbar )^{3}}}\,(\vec{n}\cdot \vec{v})\,\Theta (%
\vec{n}\cdot \vec{n}_{q})\,f(\vec{r},\vec{p},t)\right| _{{\rm surf}},
\label{1}
\end{equation}
where $\Theta (x)$ is the function which equals unity for $x>0$ and is zero
otherwise, $g_{s}=2$\ is the spin degeneracy factor of the emitted neutron,\
$\vec{n}=\vec{r}/r$ is the normal to the surface, $\vec{n}_{q}=
\vec{q}/q=\vec{p}/p=\vec{n}_{p},\,\,\,\vec{v}=
\vec{q}/m\,\,\,$, $m$ is the particle mass, and $%
E=q^{2}/2m=p^{2}/2m-V_{0}$ is the kinetic energy of the emitted neutrons. We
assume that the nucleons are confined in a spherical well potential with a
depth $V_{0}$ and time-dependent radius $R(t)$. The integrand in Eq. (\ref{1}%
) is taken at $r=R(t)$. In the phase-space approach, the distribution
function $f(\vec{r},\vec{p},t)$ in Eq. (\ref{1}) differs from the
equilibrium one due to the dynamical distortion of the Fermi surface. We use
the local-equilibrium approximation and assume the following form for the
distribution function in nuclear interior \cite{kosh04}
\begin{equation}
f(\vec{r},\vec{p},t)=\frac{1}{\exp \left\{ {\left[ (\vec{p}-m\,\vec{u}(\vec{r%
},t))^{2}/2m-\mu (t)\right] /T}\right\} +1},  \label{A1}
\end{equation}
where $\mu (t)$\ is the chemical potential, $T$\ is the temperature and $%
\vec{u}(\vec{r},t)$ is the velocity field associated with the collective
excitation of the nucleus. At small temperature $T$ such excitation
corresponds to the zero sound regime in a Fermi liquid \cite{PiNo,AbKh}
within the scaling model \cite{BoLaMa}. The distribution function of Eq. (%
\ref{A1}) implies the distortion of the Fermi surface in momentum space of
multipolarities $l=0$\ and $l=1.$ This\ is consistent with the scaling
model, where the distortions of the Fermi surface of multipolarities $l\geq 2
$ are absent due to the particular choice of the displacement field $%
\stackrel{\rightarrow }{\chi }(\vec{r},t)$ in the following form
\begin{equation}
\stackrel{\rightarrow }{\chi }(\vec{r},t)=\alpha (t)\ \overrightarrow{\nabla
}\phi (r),\quad \,\vec{u}(\vec{r},t)=\frac{\partial }{\partial t}\stackrel{%
\rightarrow }{\chi }(\vec{r},t)  \label{chi1}
\end{equation}
with
\begin{equation}
\alpha (t)=\alpha _{0}\cos (\omega t)  \label{alphat}
\end{equation}
and $\alpha _{0}$ is the (small) amplitude of oscillations. Considering only
monopole density vibrations, we put $\phi (r)=r^{2}$. The displacement field
of Eq. (\ref{chi1}) provides the disappearance of the highest
multipolarities of the Fermi surface distortion with $l\geq 2$. Thus, the
Fermi surface preserves the spherical symmetry under the assumption that the
displacement field is given by Eq. (\ref{chi1}).

Let us relate the time dependence of the chemical potential $\mu (t)$\ and
the\ radius $R(t)$ to the amplitude $\alpha (t)$. Neglecting the change in
the total number of particles $A$ during the emission process $(A\gg 1)$ we
have

\begin{equation}
A=\int \frac{g\ d\vec{r}\ d\vec{p}}{(2\pi \hbar )^{3}}\,f(\vec{r},\vec{p},t)=%
{\rm const},  \label{A}
\end{equation}
where $g=4$\ is the spin-isospin degeneracy factor. Substituting the
distribution function $\,f(\vec{r},\vec{p},t)$\ from Eq. (\ref{A1}) into Eq.
(\ref{A}) and using the boundary condition on the velocity radial component
\begin{equation}
\left. u_{r}\right| _{r=R(t)}={\frac{\partial R(t)}{\partial t},}  \label{8}
\end{equation}
we obtain the approximate relations
\begin{equation}
R(t)\approx R\left[ 1+2\,\alpha (t)+\,2\alpha ^{2}(t)\right]
,\,\,\,\,\,\,\,\,\,\mu (t)\approx \mu \left[ 1-4\,\alpha (t)+8\,\alpha
^{2}(t)\right] .  \label{10}
\end{equation}
Here, both the equilibrium values $R$ and $\mu $\ are temperature dependent\
\begin{equation}
R=R_{0}\left[ 1+\frac{\pi ^{2}}{24}\left( \frac{T}{\epsilon _{F}}\right) ^{2}%
\right] ,\quad \mu =\epsilon _{F}\left[ 1-\frac{\pi ^{2}}{12}\left( \frac{T}{%
\epsilon _{F}}\right) ^{2}\right] ,  \label{Rmu}
\end{equation}
where $R_{0}$ and $\epsilon _{F}$\ are the corresponding ground state values
of the nuclear radius and the Fermi energy, respectively.

Taking into account that the inequality $\mid m\,\vec{u}(\vec{r},t)\mid \ll
p_{F}$ is satisfied (see below Fig. 2), using Eqs. (\ref{A1}) and (\ref{chi1}%
) and keeping the lowest non-zero order of the amplitude $\alpha _{0}$, the
basic Eq. (\ref{1}) can be reduced to the following form
\begin{equation}
J_{{\rm emis}}(t)={\frac{m}{\pi ^{2}\,\hbar ^{3}}}\int_{0}^{\infty
}dE\,\,E\,\int_{0}^{1}dx\frac{x}{\exp \left\{ [E+\eta
(t)-q(E)s_{F}(t)\,x]/T\right\} +1}.  \label{14}
\end{equation}
Here, the following notations are used
\begin{equation}
s_{F}(t)=2\,{\dot{\alpha}}(t)\,R_{0}/v_{F},\,\,\,\,\,\,\,\,\eta
(t)=V_{0}-\mu (t)\approx \lambda +4\epsilon _{F}\alpha (t),\quad q(E)=2\,%
\sqrt{\epsilon _{F}\,(\epsilon _{F}+\lambda +E)},  \label{def}
\end{equation}
$\lambda =V_{0}-\epsilon _{F}$\ is the neutron separation energy, $p_{F}$ is
the Fermi momentum and $v_{F}=p_{F}/m$. The energy distribution of the
density current of the emitted particles is given from Eq. (\ref{14}) by the
following expression
\begin{equation}
{\frac{dJ_{{\rm emis}}(E,t)}{dE}}={\frac{m}{\pi ^{2}\,\hbar ^{3}}}%
\,E\,\Theta (E)\,\,\int_{0}^{1}dx\,\frac{x}{\exp \left\{ [E+\eta
(t)-q(E)s_{F}(t)\,x]/T\right\} +1}.  \label{edc}
\end{equation}

We will also consider the spectral density of emission rate $d^{2}N/dE\,dt$
derived as
\begin{equation}
{\frac{d^{2}N}{dE\,dt}}=4\,\pi \,R^{2}{\frac{dJ_{{\rm emis}}(E,t)}{dE}}.
\label{sd}
\end{equation}
The derived quantity (\ref{sd}) should be averaged over time to smear out
rapid oscillations, so the observable spectrum has the form
\begin{equation}
\overline{\frac{d^{2}N}{dE\,dt}}={\frac{1}{(2\pi /\omega )}}\,\int_{0}^{2\pi
/\omega }dt\,{\frac{d^{2}N}{dE\,dt}}.  \label{18}
\end{equation}
Here and below the bar means time-averaging.

In the case of particle emission from the vibrating cold nucleus, the
expression (\ref{edc}) reads
\begin{equation}
{\frac{dJ_{{\rm emis}}(E,t)}{dE}}={\frac{m}{\pi ^{2}\,\hbar ^{3}}}%
\,E\,\Theta (E)\,\,\int_{0}^{1}dx\,x\,\Theta \left[ q(E)s_{F}(t)\,x-\eta
(t)-E\right] .  \label{emis1}
\end{equation}
We will consider below the particle emission from the giant monopole
resonance. In this case, the amplitude $\alpha _{0}$ in Eq. (\ref{alphat})
is small enough to satisfy the condition $\eta (t)>0$ (see next section).
That means that the increase of the Fermi-sphere radius due to the
oscillations of the nuclear surface is not sufficient to cause the emission
of particles, i.e., the particle emission appears only because of the
dynamic Fermi-surface shift. In this case, Eq. (\ref{edc}) takes the
following explicit form
\begin{equation}
{\frac{dJ_{{\rm emis}}(E,t)}{dE}}={\frac{m}{2\,\pi ^{2}\,\hbar ^{3}}}%
\,E\,\Theta (E)\,\Theta \lbrack s_{F}(t)]\,\Theta \lbrack 1-\xi (t,E)]\left[
1-\xi ^{2}(t,E)\right] ,  \label{emis2}
\end{equation}
where
\begin{equation}
\xi (t,E)={\frac{\eta (t)+E}{s_{F}(t)\,q(E)}}.  \label{zeta}
\end{equation}

Eq. (\ref{emis2}) allows us to evaluate the maximal value of the
kinetic energy, $E_{{\rm max}}$, of the emitted particles. The
result reads
\begin{equation}
E_{{\rm max}}\approx -\lambda +{\frac{10}{3}}\,{\frac{E^{\ast }}{A}}\,\left[
1+\sqrt{1+{\frac{3}{5}}\,A\,{\frac{\epsilon _{F}}{E^{\ast }}}+{\frac{12}{5}}%
\,{\frac{A}{(k_{F}\,R_{0})^{2}}}\,\left( {\frac{\epsilon _{F}}{E^{\ast }}}%
\right) ^{3}}\right] ,  \label{19}
\end{equation}
where $k_{F}=p_{F}/\hbar $ and $E^{\ast }$ is the energy of the collective
excitation.

\section{\protect\bigskip Numerical calculations and discussion}

We will apply Eqs. (\ref{edc}) and (\ref{emis2}) to the particle emission
from the isoscalar giant monopole resonance ({\rm ISGMR}). The
eigenfrequency $\omega $\ of the ISGMR in Eq. (\ref{alphat}) can be found
from the classical derivation $\omega =\sqrt{C/B},$\ where $C$\ and $B$\ are
the stiffness and the mass coefficient, respectively. Both transport
coefficients $C$\ and\ $B,$ can be obtained evaluating the collective
potential, $E_{{\rm pot}},$ and kinetic, $E_{{\rm kin}},$\ energy of the
nucleus within the scaling approximation (\ref{chi1}). Using Eq. (\ref{chi1}%
), we obtain
\begin{equation}
E_{{\rm kin}}={\frac{1}{2}}\,m\,\int d^{3}r\ \rho _{{\rm eq}}\,u^{2}\approx {%
\frac{6}{5}}\,A\,m\,R_{0}^{2}\ \,{\dot{\alpha}}^{2}(t)=\frac{1}{2}B\ {\dot{%
\alpha}}^{2}(t)  \label{ekin}
\end{equation}
and
\begin{equation}
B=(12/5)AmR_{0}^{2}.  \label{b1}
\end{equation}
Here, $\rho _{{\rm eq}}$\ is the equilibrium particle density.

The collective potential energy $E_{{\rm pot}}$\ is derived as
\begin{equation}
E_{{\rm pot}}=\frac{1}{2}\int d^{3}r\left( \frac{\delta ^{2}{\cal E}}{\delta
\rho ^{2}}\right) _{{\rm eq}}(\delta \rho )^{2},  \label{epot1}
\end{equation}
where $\epsilon $ is the particle energy density\ which is related to the
static incompressibility $K$\ by
\begin{equation}
K=9\left( \rho \frac{\delta ^{2}{\cal E}}{\delta \rho ^{2}}\right) _{{\rm eq}%
}.  \label{k1}
\end{equation}
Using the continuity equation, $\delta \rho =-\stackrel{\rightarrow }{\nabla
}\cdot $\ $\rho _{{\rm eq}}\stackrel{\rightarrow }{\chi },$ we obtain from
Eqs. (\ref{chi1}) and (\ref{epot1})
\begin{equation}
E_{{\rm pot}}\approx 2KA\alpha ^{2}(t)=\frac{1}{2}C\alpha ^{2}(t),\quad C=4KA
\label{epot2}
\end{equation}
and
\begin{equation}
\omega =\sqrt{\frac{K}{m\ \left\langle r^{2}\right\rangle }},  \label{om1}
\end{equation}
where $\left\langle r^{2}\right\rangle =(3/5)R_{0}^{2}$\ is the mean square
radius of the nucleus. We point out that in the case of finite nuclei the
incompressibility $K$\ is $A$-dependent.

Using Eqs. (\ref{alphat}) and (\ref{ekin}), one can evaluate the amplitude $%
\alpha _{0}$ of the Fermi-surface oscillations. It can be done using the
expression for the collective kinetic energy $E_{{\rm kin}}$ of the monopole
vibrations. Averaging Eq. (\ref{ekin}) over time within the period of
oscillations $2\pi /\omega $ and using the virial theorem we find
\begin{equation}
\alpha _{0}^{2}={\frac{5}{3}}\,{\frac{1}{A}}\,{\frac{1}{(k_{F}\,R_{0})^{2}}}%
\,{\frac{\epsilon _{F}\ E^{\ast }}{(\hbar \omega )^{2}}}.  \label{13}
\end{equation}
In the case of particle emission from the GMR, we put $E^{\ast }=\hbar
\omega =E_{GMR},$ where $E_{GMR}$\ is the eigenenergy of the ISGMR. In this
work we adopt the values of $R_{0}=1.12\cdot \ A^{1/3}\,{\rm fm,}$ $\epsilon
_{F}=37$ ${\rm MeV}$, $k_{F}=1.36$ fm$^{-1},$ $E_{GMR}=82\cdot A^{-1/3}\,%
{\rm MeV}$\ and $\lambda =7$ ${\rm MeV}$.

\begin{figure}[tbp]
\includegraphics[width=4.0 in,height=4.0 in]{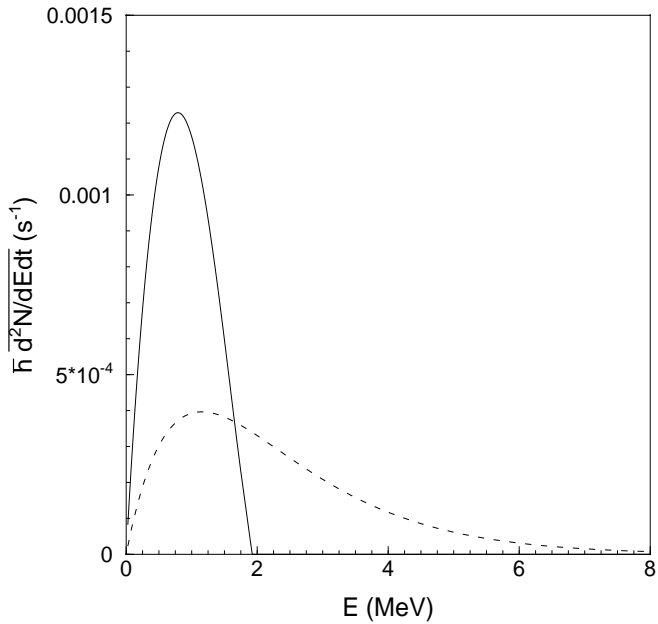}
\caption{The spectral density of emission rate $\hbar \overline{d^{2}N/dE\,dt%
}$ for splashing of neutrons from the giant monopole resonance
(GMR) in cold nucleus (solid curve) and for particle evaporation
from thermal equilibrated nucleus (dashed curve) versus particle
kinetic energy $E$ of emitted
neutrons. We have used the values of GMR energy $E_{GMR}=82\cdot A^{-1/3}\,%
{\rm MeV}$, thermal excitation energy $E_{T}=E_{GMR}$ and mass
number $A=90$.}
\end{figure}

In{\rm \ Fig. 1} the dependence of $\overline{d^{2}N/dE\,dt}$, obtained from
Eqs. (\ref{emis1}), (\ref{sd}) and (\ref{18}), on the kinetic energy of the
emitted particle $E$ is shown (solid line) for the nucleus with $A=90$ and $%
T=0$. We will compare this result of particle emission from the ISGMR in the
cold nucleus with the one for the evaporation of neutrons from the thermal
equilibrated nucleus with the thermal excitation energy $E_{T}=E_{GMR}$. For
the case of thermal evaporation of particles, we take\- $\mu (t)=\lambda $
and $\vec{u}(\vec{r},t)=0$\ in the distribution function $f(\vec{r},\vec{p}%
,t)$\ of Eq. (\ref{A1}).\ \ Using Eqs. (\ref{edc}) and (\ref{sd}), we
obtain, see also Ref. \cite{FrLy},
\begin{equation}
\left. {\frac{d^{2}N}{dE\,dt}}\right| _{{\rm evap}}={\frac{(k_{F}R_{0})^{2}}{%
\pi \,\hbar \,\epsilon _{F}}}\,E\,\Theta (E)\,{\frac{1}{{1+{\rm exp}[(E+V}%
_{0}-\lambda {)/T]}}}.  \label{20}
\end{equation}
The dashed curve in {\rm Fig. 1} represents the spectral density of
evaporation rate (\ref{20}) from a heated nucleus with the temperature $T=%
\sqrt{8\,E_{T}/A}$. As seen from {\rm Fig. 1}, the thermal spectrum (dashed
line) is much broader than the cold emission spectrum (solid line) and has
its maximum at a larger energy.

{\rm Fig. 1} shows that the kinetic energy of the emitted particle from a
cold nucleus is much smaller than the Fermi energy $\epsilon _{F}$, so in (%
\ref{zeta}) we can neglect the dependence of $q(E)$ upon $E$. Integrating (%
\ref{emis2}) with $q(E)=q_{0}=2\,\sqrt{\epsilon _{F}\,(\epsilon _{F}+\lambda
)}$ over energy $E$, we obtain an analytical expression for the cold
particle emission rate $dN/dt$\ as
\[
\frac{dN}{dt}=4\pi R_{0}^{2}\ J_{{\rm emis}}(t)
\]
\begin{equation}
=\frac{(k_{F}R_{0})^{2}}{4\pi \hbar \epsilon _{F}}[s_{F}(t)\
q_{0}]^{2}\,\Theta \,[s_{F}(t)]\,\Theta \lbrack 1-\beta (t)]\left\{ 1-{\frac{%
8}{3}}\,\beta (t)+2\,\beta ^{2}(t)-{\frac{1}{3}}\,\beta ^{4}(t)\right\} ,
\label{er}
\end{equation}
where $\beta (t)=\eta (t)/q_{0}s_{F}(t)$. {\rm Fig. 2} shows the
emission rate $dN/dt$ of Eq. (\ref{er}) for the interval of time
$\Delta t\leq 2\pi /\omega $. As seen from Fig. 2 the particle
emission from ISGMR occurs as a short time splashing.

\begin{figure}[tbp]
\includegraphics[width=4.0 in,height=4.0 in]{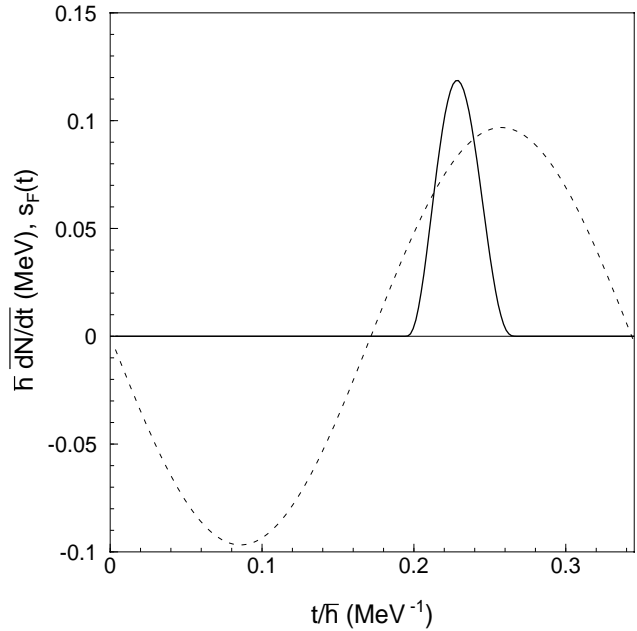}
\caption{The emission rate $dN/dt$ of neutrons from the giant
monopole resonance in cold nucleus with $A=90$ (solid line). The
dashed line shows the time dependence of the dimensionless
parameter, $s_{F}(t)$, of the Fermi surface distortion, see Eq.
(\ref{def}).}
\end{figure}

The particle emission rate $dN/dt$ can be used to derive the life
time, $\tau _{{\rm cold}},$ with respect to the splashing of
neutron from the ISGMR.\ The quantity (\ref{er}) must be averaged
over time as in (\ref{18}), resulting in
\begin{equation}
{\frac{1}{\tau _{{\rm
splash}}}}=\overline{\frac{dN}{dt}}={\frac{1}{(2\pi /\omega
)}}\,\int_{0}^{2\pi /\omega }dt\,{\frac{dN}{dt}}.  \label{tau}
\end{equation}
For the nucleus with $A=90$, one obtains from Eqs. (\ref{er}) and (\ref{tau}%
) that $\tau _{{\rm splash}}=6.0\cdot 10^{-20}$ s. The analogous
quantity for
particle evaporation, calculated by means of (\ref{20}), is $\tau _{{\rm evap%
}}=5.3\cdot 10^{-19}$ s. The calculations of $\tau _{{\rm
splash}}$\ can be improved if the higher multipolarities of the
Fermi surface distortion, at least the quadrupole, are taken into
account, see Appendix.

\section{Conclusions\newline
}

\bigskip

By use of phase space approach to the particle emission\ from the
nucleus, we have connected the particle emission rate to the
dynamical distortion\ of the Fermi surface of the cold vibrating
nucleus. The particle emission occurs here as a classical
splashing effect from the vibrating liquid drop. We have shown
that cold particle emission (splashing) may result in the
deviation of the observed spectra from the usual statistical ones
in the small energy region, see Fig. 1. In particular, for the
cold particle emission, the emission rate, $(dN/dt)_{{\rm
splash}},$ (averaged over time) is significantly large then the
corresponding value, $(dN/dt)_{{\rm evap}}$,
for the particle evaporation (for the ISGMR $(dN/dt)_{{\rm splash}}$ exceeds $%
(dN/dt)_{{\rm evap}}$ by factor about 10).

The general conclusion of this paper is that the collective motion in a
finite nuclear Fermi liquid is accompanied by direct (non-statistical)
emission of nucleons via the dynamical distortion of the Fermi surface. This
mechanism of cold particle emission\ can be also applied to the large
amplitude motion like a descent of the nucleus from the fission barrier to
the scission point or to the first stage of the heavy ion collision.

\bigskip

\section{ Acknowledgments}

The author thanks H. Hofmann and R. Hilton for valuable
suggestions and the Physics Department of the Technische
Universit\"{a}t M\"{u}nchen for the nice hospitality. Financial
support by the Deutsche Forschungsgemeinschaft under contract 436
UKR 113/66/0-1 is gratefully acknowledged.

\newpage

\setcounter{equation}{0} \renewcommand{\theequation}{A\arabic{equation}}
\appendix

\begin{center}
{\bf APPENDIX A}
\end{center}



We will generalize the local-equilibrium approximation of Eq. (\ref{A1}) and
assume the following non-equilibrium form for the distribution function in
nuclear interior at $T=0$, see Ref. \cite{kosh04},
\begin{equation}
f(\vec{r},\vec{p},t)=\Theta \left[ {\frac{1}{2m}}p_{F}^{2}(\vec{r},t)-{\frac{%
1}{2m}}[\vec{p}-m\,\vec{u}(\vec{r},t)]^{2}-\nu (\vec{r},\vec{p},t)\right] ,
\label{xx1}
\end{equation}
where $p_{F}(\vec{r},t)$\ is the Fermi momentum and $\nu (\vec{r},\vec{p},t)$
is associated with the distortion of the Fermi-surface in the momentum
space.\

To describe the collective excitations with the multipolarity $L$ in the
spherical well potential we will consider $\nu (\vec{r},\vec{p},t)$ as a
superposition of plane wave to create a state with a good angular moment $L$%
, see also \cite{JeJa,MaKoHoSh}. Namely,
\begin{equation}
\nu (\vec{r},\vec{p},t)=\int d\Omega _{k}\,\,Y_{L0}(\vec{n}_{k})\,F(\vec{n}%
_{p}\cdot \vec{n}_{k})\,{\rm exp}[i(\vec{k}\cdot \vec{r}-\omega t)]+{\rm c.c.%
},  \label{A2}
\end{equation}
where
\begin{equation}
F(\vec{n}_{p}\cdot \vec{n}_{k})=\sum_{l\neq 1}\nu _{l}^{(0)}\,Y_{l0}(\vec{n}%
_{p}\cdot \vec{n}_{k})  \label{A3}
\end{equation}
and $\nu _{l}(t)=\nu _{l}^{(0)}\,(e^{-i\,\omega \,t}+{\rm c.c.})$ is the
amplitude of the Fermi-surface distortion of multipolarity $l$ in the
momentum space. We point out that the condition $l\neq 1$ in Eq. (\ref{A3})
is because the shift of the Fermi surface was already extracted in the
distribution function $f(\vec{r},\vec{p},t)$, taken in the form of Eq. (\ref
{xx1}). We will follow the nuclear fluid dynamic approach, and take into
account the dynamic Fermi-surface distortion up to multipolarity $l=2$ in
the expansion of Eq. (\ref{A3}). In this case, the amplitudes $\nu _{0}$ and
$\nu _{2}$ are related to each other as $\nu _{2}=\sqrt{4/5}\,\nu _{0}$, see
Ref. \cite{YuHo}. Finally, putting $Y_{L=0,0}=1/\sqrt{4\pi }$ in Eq. (\ref
{A2}), we obtain the distortion amplitude $\nu (\vec{r},\vec{p},t)$ for
monopole mode in the following form
\begin{equation}
\nu (\vec{r},\vec{p},t)=\nu _{0}\,\left[ j_{0}(kr)-2\,j_{2}(kr)\,P_{2}(x)%
\right] \,e^{-i\,\omega \,t}+{\rm c.c.},\,\,\,\,\,\,{\rm for}\,\,\,\,\,\,L=0,
\label{A4}
\end{equation}
where $j_{l}(kr)$ is the spherical Bessel function, $P_{l}(x)$ is the
Legendre polynomial and $x=\vec{n}\cdot \vec{n}_{p}$.

The amplitude $\nu _{0}$ is related to the bulk density variation $\delta
\rho (\vec{r},t)$. Namely, using Eqs. (\ref{xx1}) and (\ref{A4}), one
obtains the following expression in the nuclear interior
\begin{equation}
\delta \rho (\vec{r},t)=\int {\frac{d\vec{p}}{(2\pi \hbar )^{3}}}\,f(\vec{r},%
\vec{p},t)-\rho _{{\rm eq}}(\vec{r})=\nu
_{0}^{(0)}\,N_{0}\,j_{0}(kr)\,e^{-i\,\omega \,t}+{\rm c.c.}%
,\,\,\,\,\,(r<R,\,\,L=0),  \label{A5}
\end{equation}
where $N_{0}=2\,m\,p_{F}/\pi ^{2}\hbar ^{3}$ is the density of single
particle states on the Fermi-surface. The velocity field $\vec{u}(\vec{r},t)$
can be evaluated from the continuity equation
\begin{equation}
{\frac{\partial }{\partial t}}\delta \rho (\vec{r},t)=-\vec{\nabla}\cdot
\rho _{{\rm eq}}\,\vec{u}(\vec{r},t).  \label{A6}
\end{equation}
Using Eqs. (\ref{A5}) and (\ref{A6}), one obtains
\begin{equation}
\vec{u}(\vec{r},t)=2\,\nu _{0}\,N_{0}\,\omega \,{\frac{j_{1}(kr)}{\rho
_{0}\,kr}}\vec{r}\,\sin (\omega t),\,\,\,\,\,\,(L=0).  \label{A7}
\end{equation}
The wave number $k$ in Eqs. (\ref{A4}), (\ref{A5}) and (\ref{A7}) is derived
from the boundary condition. The boundary condition can be taken as a
condition for the balance, at the nuclear surface, between the compressional
pressure and the surface tension pressure, see \cite{BoMo2,KoKoSh}. Taking
into account the consistent change of the compressional pressure due to the
Fermi-surface distortion, the boundary condition reads \cite{KoKoSh}
\begin{equation}
z_{n}\,j_{0}(z_{n})-(f_{\sigma }+f_{\mu })\,j_{1}(z_{n})=0.  \label{A8}
\end{equation}
Here, $z=kR$,
\begin{equation}
f_{\sigma }={\frac{18\,\sigma }{\rho _{0}\,R_{0}\,K^{\prime }}}%
,\,\,\,\,\,\,f_{\mu }={\frac{36\,\mu }{K^{\prime }}},  \label{ff}
\end{equation}
where $\sigma $ is the surface tension coefficient, $K^{\prime }$ is the
dynamic incompressibility given by
\begin{equation}
K^{\prime }=K+K_{\mu }  \label{eq1}
\end{equation}
and $K$ is the commonly used {\it static} incompressibility
\begin{equation}
K=R^{2}\,{\frac{\delta ^{2}{\cal E}/A}{\delta R^{2}}}\Big\vert_{R=R_{0}}.
\label{eq2}
\end{equation}
The additional contribution, $K_{\mu }$, to the incompressibility $K^{\prime
}$ in Eq. (\ref{eq1}) is due to the {\it dynamic} Fermi-surface distortion
effect. The quantity $K_{\mu }$ can be evaluated in a general case of
arbitrary multipolarity of the Fermi-surface distortion and is given by \cite
{kosh04,kmpl,KoKoSh}
\begin{equation}
K_{\mu }=12\,\epsilon _{F}\,{\frac{\Omega _{20}(s)}{\Omega _{00}(s)}}%
\,\,\,\,\,\,{\rm with}\,\,\,\,\,\Omega _{l0}(s)={\frac{1}{2}}%
\,\int_{-1}^{1}dx\,\,{\frac{x\,P_{l}(x)\,P_{0}(x)}{{x-s}}},  \label{kF}
\end{equation}
where the dimensionless zero-sound velocity $s=\hbar \omega /p_{F}k$ is
found from the Landau's dispersion relation \cite{AbKh,lp}. In the case of
an isotropic Landau's interaction amplitude, i.e., $F_{0}\neq 0,\quad
F_{l\neq 0}=0$, the dispersion relation reads
\begin{equation}
\Omega _{00}(s)=-\frac{1}{F_{0}}.{\rm \quad }  \label{disp2}
\end{equation}
For realistic nuclear forces, $F_{0}\sim 0$, Eqs. (\ref{kF}) and (\ref{disp2}%
) lead to large renormalization of the incompressibility with $K_{\mu
}\approx 2K$. The giant monopole resonance (GMR) corresponds to the lowest
solution, $z_{0}>1$, to the secular equation (\ref{A8}).

Using Eqs. (\ref{xx1}), (\ref{A4}) and (\ref{A7}), the flux density $J_{{\rm %
emis}}(t)$ of Eq. (\ref{1}) can be evaluated beyond the scaling
approximation used in Sect. II. \ \


\end{document}